\documentclass[twocolumn,aps,prl,showpacs,superscriptaddress,floatfix]{revtex4}
\usepackage{dcolumn}
\usepackage{epsf}
\usepackage{bm}
\usepackage{times}

\usepackage{graphicx}

\newcommand{\addrGaithersburg}{National Institute of Standards and Technology, 
Mail Stop 8401, Gaithersburg, MD 20899-8401, USA}
\newcommand{\addrHeidelberg}{Max--Planck--Institut f\"ur Kernphysik,
Saupfercheckweg 1, 69117 Heidelberg, Germany}
\newcommand{\addrParis}{Laboratoire Kastler Brossel, \'Ecole Normale Sup\'erieure et
Universit\'{e} Pierre et Marie Curie, Case 74\\
4, pl.\ Jussieu, 75005 Paris, France}
\begin{document}

\title{Precise calculation of transition frequencies of hydrogen and deuterium
 \\ based on a least-squares analysis}
%\title{Calculation of energy levels of hydrogen and deuterium}

\author{Ulrich D.~Jentschura}
\affiliation{\addrGaithersburg}
\affiliation{\addrHeidelberg}

\author{Svetlana Kotochigova}
\affiliation{\addrGaithersburg}

\author{Eric-Olivier Le~Bigot}
\affiliation{\addrGaithersburg}
\affiliation{\addrParis}

\author{Peter J.~Mohr}
\affiliation{\addrGaithersburg}

\author{Barry N. Taylor}
\affiliation{\addrGaithersburg}

\date{\today}

\begin{abstract}
We combine a limited number of accurately measured transition frequencies in hydrogen and 
deuterium, recent quantum electrodynamics (QED) calculations, and, as an 
essential additional ingredient, a generalized least-squares analysis,
to obtain precise and optimal predictions for hydrogen and
deuterium transition frequencies.
Some of the predicted transition frequencies have
relative uncertainties more than an order of magnitude
smaller than that of the $g$-factor
of the electron, which was previously the most accurate
prediction of QED.
\end{abstract}
%
%    48 x 11 + 4  = 532 characters in abstract
%
%
\maketitle
%
%%%%%%%%%%%%%%%%%%%%%%%%%%%%%%%%%%%%%%%%%%%%%%%%%%%%%%%%%%%%%%%%%%%%%

In the past decade there have been significant advances in both the
measurement and theory of transition frequencies in hydrogen
and deuterium.  With the advent of frequency-comb metrology,
the progress in measurements has accelerated to the extent that the
uncertainty in the experimental value of the 1S--2S frequency in 
hydrogen has decreased by three orders of magnitude in about
ten years \cite{2000034}.  Many other precision measurements of
transitions in hydrogen and deuterium with principal quantum number 
$n$ up to 12 have been made and are listed in Table~\ref{tab:rydfreq},
together with the 1S--2S frequency
\cite{2000034,1995159,1998002,1997001,1999072,1996001,1995138,1994090,
1986003,1979001}.
In addition, precise measurements have been made involving states 
of hydrogen with $n=27$ to 30 \cite{th02dv}.
Further advances may be anticipated from a number of groups currently
working to improve measurement accuracy
\cite{pc04th,pc04fn,pc04fm,pc04dk}.

%%%%%%%%%%%%%%%%%%%%%%%%%%%%%%%%%%%%%%%%%%%%%%%%%%%%%%%%%%%%%%%%%%%%%

\def\vsp{\vbox to  10pt{}}
 \begin{table*}
 \caption{
 Transition frequencies 
 in hydrogen $\nu_{\rm H}$ and in deuterium $\nu_{\rm D}$
used in the 2002 CODATA least-squares adjustment of the 
values of the fundamental constants and the calculated values.
Hyperfine effects are not included in these values.
 }
 \label{tab:rydfreq}
 \begin{tabular}{l@{\quad}l@{\quad}l@{\quad}l}
 \toprule
 \noalign{\vbox to 5 pt {}}
  Experiment  
   & \hbox to 23 pt {} Frequency interval(s) & Reported value
  & Calculated value \\ 
  & & 
 \hbox to 10pt{} $\nu$/kHz & \hbox to 10pt{} $\nu$/kHz \\
 \noalign{\vbox to 5 pt {}}
 \colrule
 \noalign{\vbox to 5 pt {}}
 \citet{2000034} 
 & $\nu_{\rm H}({\rm 1S_{1/2}}-{\rm 2S_{1/2}})$ 
 & $ 2\,466\,061\,413\,187.103(46)$ 
 & $ 2\,466\,061\,413\,187.103(46)$ \\ 
 \vsp\citet{1995159}  
 & $\nu_{\rm H}({\rm 2S_{1/2}}-{\rm 4S_{1/2}})
 - {1\over4}\nu_{\rm H}({\rm 1S_{1/2}}-{\rm 2S_{1/2}})$
 & $ 4\,797\,338(10)$ 
 & $ 4\,797\,331.8(2.0)$ \\
 & $\nu_{\rm H}({\rm 2S_{1/2}}-{\rm 4D_{5/2}})
 - {1\over4}\nu_{\rm H}({\rm 1S_{1/2}}-{\rm 2S_{1/2}})$
  & $ 6\,490\,144(24)$ 
  & $ 6\,490\,129.9(1.7)$ \\
 & $\nu_{\rm D}({\rm 2S_{1/2}}-{\rm 4S_{1/2}})
 - {1\over4}\nu_{\rm D}({\rm 1S_{1/2}}-{\rm 2S_{1/2}})$
 & $ 4\,801\,693(20)$ 
 & $ 4\,801\,710.2(2.0)$ \\
 & $\nu_{\rm D}({\rm 2S_{1/2}}-{\rm 4D_{5/2}}) 
 - {1\over4}\nu_{\rm D}({\rm 1S_{1/2}}-{\rm 2S_{1/2}})$
  & $ 6\,494\,841(41)$ 
  & $ 6\,494\,831.5(1.7)$ \\
 \vsp\citet{1998002} 
 & $\nu_{\rm D}({\rm 1S_{1/2}} -{\rm 2S_{1/2}}) 
 - \nu_{\rm H}({\rm 1S_{1/2}} - {\rm 2S_{1/2}})$ 
 & $ 670\,994\,334.64(15)$ 
 & $ 670\,994\,334.64(15)$  \\
 
 \vsp\citet{1997001} 
 & $\nu_{\rm H}({\rm 2S_{1/2}}-{\rm 8S_{1/2}})$ & 
 $ 770\,649\,350\,012.0(8.6)$ 
 & $ 770\,649\,350\,016.1(2.8)$  \\ 
 & $\nu_{\rm H}({\rm 2S_{1/2}}-{\rm 8D_{3/2}})$ & 
 $ 770\,649\,504\,450.0(8.3)$ 
 & $ 770\,649\,504\,449.1(2.8)$  \\ 
 & $\nu_{\rm H}({\rm 2S_{1/2}}-{\rm 8D_{5/2}})$ & 
 $ 770\,649\,561\,584.2(6.4)$ 
 & $ 770\,649\,561\,578.2(2.8)$  \\ 
 & $\nu_{\rm D}({\rm 2S_{1/2}}-{\rm 8S_{1/2}})$ & 
 $ 770\,859\,041\,245.7(6.9)$ 
 & $ 770\,859\,041\,242.6(2.8)$  \\ 
 & $\nu_{\rm D}({\rm 2S_{1/2}}-{\rm 8D_{3/2}})$ & 
 $ 770\,859\,195\,701.8(6.3)$ 
 & $ 770\,859\,195\,700.3(2.8)$  \\ 
 & $\nu_{\rm D}({\rm 2S_{1/2}}-{\rm 8D_{5/2}})$ & 
 $ 770\,859\,252\,849.5(5.9)$ 
 & $ 770\,859\,252\,845.1(2.8)$  \\ 
 
 \vsp\citet{1999072} 
 & $\nu_{\rm H}({\rm 2S_{1/2}}-{\rm 12D_{3/2}})$ & 
 $ 799\,191\,710\,472.7(9.4)$ 
 & $ 799\,191\,710\,481.9(3.0)$  \\ 
 & $\nu_{\rm H}({\rm 2S_{1/2}}-{\rm 12D_{5/2}})$ & 
 $ 799\,191\,727\,403.7(7.0)$ 
 & $ 799\,191\,727\,409.1(3.0)$  \\ 
 & $\nu_{\rm D}({\rm 2S_{1/2}}-{\rm 12D_{3/2}})$ & 
 $ 799\,409\,168\,038.0(8.6)$ 
 & $ 799\,409\,168\,041.7(3.0)$  \\ 
 & $\nu_{\rm D}({\rm 2S_{1/2}}-{\rm 12D_{5/2}})$ & 
 $ 799\,409\,184\,966.8(6.8)$ 
 & $ 799\,409\,184\,973.4(3.0)$  \\ 
 
 \vsp \citet{1996001} 
 & $\nu_{\rm H}({\rm 2S_{1/2}}-{\rm 6S_{1/2}})
 - {1\over4}\nu_{\rm H}({\rm 1S_{1/2}}-{\rm 3S_{1/2}})$
 & $ 4\,197\,604(21)$ 
 & $ 4\,197\,600.3(2.2)$ \\ 
 & $\nu_{\rm H}({\rm 2S_{1/2}}-{\rm 6D_{5/2}})
 - {1\over4}\nu_{\rm H}({\rm 1S_{1/2}}-{\rm 3S_{1/2}})$
 & $ 4\,699\,099(10)$ 
 & $ 4\,699\,105.4(2.2)$ \\ 
 
 \vsp\citet{1995138} 
 & $\nu_{\rm H}({\rm 2S_{1/2}}-{\rm 4P_{1/2}})
 - {1\over4}\nu_{\rm H}({\rm 1S_{1/2}}-{\rm 2S_{1/2}})$
 & $ 4\,664\,269(15)$ 
 & $ 4\,664\,254.3(1.7)$ \\ 
 & $\nu_{\rm H}({\rm 2S_{1/2}}-{\rm 4P_{3/2}})
 - {1\over4}\nu_{\rm H}({\rm 1S_{1/2}}-{\rm 2S_{1/2}})$
 & $ 6\,035\,373(10)$ 
 & $ 6\,035\,384.1(1.7)$ \\ 
 
 \vsp\citet{1994090} 
 & $\nu_{\rm H}({\rm 2S_{1/2}}-{\rm 2P_{3/2}})$
 & $ 9\,911\,200(12)$ 
 & $ 9\,911\,197.6(2.4)$ \\ 
 
 \vsp\citet{1986003} 
 & $\nu_{\rm H}({\rm 2P_{1/2}}-{\rm 2S_{1/2}})$
 & $ 1\,057\,845.0(9.0)$ 
 & $ 1\,057\,844.0(2.4)$ \\ 
 
 \vsp\citet{1979001} 
 & $\nu_{\rm H}({\rm 2P_{1/2}}-{\rm 2S_{1/2}})$
 & $ 1\,057\,862(20)$ 
 & $ 1\,057\,844.0(2.4)$ \\
 \botrule
 \end{tabular}
 \end{table*}

%%%%%%%%%%%%%%%%%%%%%%%%%%%%%%%%%%%%%%%%%%%%%%%%%%%%%%%%%%%%%%%%%%%%%

On the theoretical side, there has been 
progress in the accuracy of quantum electrodynamics (QED) 
calculations which has led to a significant reduction of
the uncertainty of both the one-photon 
\cite{1999001,2001072,2003108,2003187,jmpra04}
and two-photon 
\cite{1996012,1996192,2001059,2002155,2003160,2003118}
contributions.
In addition to this progress, the application of the least-squares method
described here makes it possible in some cases to predict transition 
frequencies with a relative uncertainty that is more than an order of 
magnitude smaller than the relative uncertainty in the Rydberg constant,
which is $6.6\times10^{-12}$.

It is of interest to have accurate calculations of hydrogen and deuterium
transition frequencies for comparison to existing and new experimental values, 
for both frequency standards applications, and as tests of QED.
Also, with sufficiently accurate theory available, it would be possible
to consider redefining the SI second in terms of an assigned value of the 
Rydberg constant.  Although this would entail about three orders of
magnitude improvement in the theory, the recent rate of progress suggests
that it may be within reach.

In this letter, we give theoretical values for a number of transition 
frequencies that are meant to be best values consistent with currently available 
experimental and theoretical information.  
Such calculated values for the transitions in Table~\ref{tab:rydfreq} 
follow from the results of the 2002 CODATA least-squares adjustment of the 
fundamental constants 
\cite{codata2002,codata1998,1934003},
and are listed in that table.  
Of course, the calculated value of the 1S--2S
transition in hydrogen should not be viewed as a theoretical prediction,
because the least-squares adjustment determines values of
the adjusted constants (variables or unknowns of the adjustment)
so that the calculated frequency is essentially 
equal to the very precise measured value.
The number in parentheses with each value is the standard
uncertainty (estimated standard deviation) in the last two figures.  
Hyperfine structure effects are not included in the quoted numbers.

For hydrogen or deuterium transition frequencies 
not included in the 2002 adjustment, we proceed as follows.
The energy level $E_i$ of state $i$ can 
be written as a function of the fundamental constants and an
additional adjusted constant $\delta_i$ which takes into account the uncertainty 
in the theory \cite{1990020,2001057,codata2002}.  
For example, for the case in which $i$ is a state of hydrogen, we have
\begin{eqnarray}
E_i &=& H_i\big[R_\infty,\alpha,A_{\rm r}({\rm e}), A_{\rm r}({\rm p}),
R_{\rm p}\big] + \delta_i ,
\label{eq:el}
\end{eqnarray}
where the constants that appear as arguments of the function $H_i$ are listed in
Table~\ref{tab:codata2004}.  Because the values of the constants in Eq.~(\ref{eq:el}),
including $\delta_i$, result from a least-squares adjustment, they are correlated, 
particularly those for $R_\infty$ and $R_{\rm p}$,
which have a correlation coefficient of 0.996.  
The uncertainty of the calculated value for the 1S--2S frequency in hydrogen
is increased by a factor of about 500 if such correlations are neglected.
The function $H_i$ also depends implicitly on $c$ and the Planck constant $h$.
However, these constants are not displayed as variables, because $c$ is a fixed number, 
and the frequencies $(E_i-E_{i^\prime})/h$ are essentially independent of $h$.
Levels in deuterium are given as similar functions
with p replaced by d.  

%%%%%%%%%%%%%%%%%%%%%%%%%%%%%%%%%%%%%%%%%%%%%%%%%%%%%%%%%%%%%%%%%%%%%
\def\vs{\vbox to 4pt{}}
\begin{table}[b]
\vspace{-15 pt} 
\caption{
The CODATA 2002 values of the constants used
in the evaluation of the spectrum of hydrogen and deuterium.}
\label{tab:codata2004}
\begin{tabular}{l@{}l}
\toprule
\noalign{\vbox to 2 pt {}}
  Constant   &  Value {} \\ 
\colrule
\noalign{\vbox to 2 pt {}}

  Speed of light \vs
& $ c \!=\! 299\,792\,458$~m~s$^{-1}$  \\ 

  Rydberg constant \vs
& $ R_\infty \!\!=\! 10\,973\,731.568\,525(73)$~m$^{-1}$  \\ 

  Fine-structure constant \vs
& $ \alpha \!=\! 1/137.035\,999\,11(46)$ \\ 

  Electron relative atomic mass \vs
& $ A_{\rm r}({\rm e}) \!=\! 5.485\,799\,0945(24)\!\!\times\!\!10^{-4}$  \\ 

  Proton relative atomic mass \vs
& $ A_{\rm r}({\rm p}) \!=\! 1.007\,276\,466\,88(13)$  \\ 

  Deuteron relative atomic mass \vs
& $ A_{\rm r}({\rm d}) \!=\! 2.013\,553\,212\,70(35)$  \\ 

  Proton rms charge radius \vs
& $ R_{\rm p} \!=\! 0.8750(68)$~fm  \\ 

  Deuteron rms charge radius \vs
& $ R_{\rm d} \!=\! 2.1394(28)$~fm  \\ 

\botrule
\end{tabular}
\end{table}

%%%%%%%%%%%%%%%%%%%%%%%%%%%%%%%%%%%%%%%%%%%%%%%%%%%%%%%%%%%%%%%%%%%%%

The theory included in the function $H_i$ in Eq.~(\ref{eq:el}) is described in
detail in Appendix A of Ref.~\cite{codata2002}, which provides a review of the
relevant calculations.  Much of that information is in the form of equations
that are valid for any state, with the exception of tables of data that only
have entries for the levels included in the 2002 CODATA adjustment.  Enlarged
versions of those tables with data for all states with $n\le200$ are
available on the NIST Physics Laboratory Web site at physics.nist.gov/hdel.
Estimates of the theoretical uncertainties of the function $H_i$, represented by
the constant $\delta_i$ in Eq.~(\ref{eq:el}), are also given in Appendix A of
Ref.~\cite{codata2002}.  The a priori 
estimated value of $\delta_i$ is $\delta_i({\rm th})=0$,
because the theoretical expression for the levels includes all known 
contributions.  However, the estimated uncertainty $u[\delta_i({\rm th})]$
is not zero, and there are significant
covariances between the various $\delta$s that take into account the expected
patterns in the uncertainties.  For example, for S states there are 
components of uncertainty with the functional form $C/n^3$, where $C$ is a common 
unknown constant, and there are components of uncertainty
common to hydrogen and deuterium levels with the same quantum numbers.
The theoretical uncertainties and covariances are included in the least-squares
adjustment as input data for the adjusted variables $\delta_i$.

A generalized least-squares adjustment is formulated here along the lines
described in Refs.~\cite{1934003} and \cite{codata1998}.
Symbols that refer to data used in the 2002 CODATA adjustment of the constants 
are also defined in Ref.~\cite{codata1998}.
New energy levels $E_l$ to be determined are added to the adjustment, along with the
corresponding theoretical expressions of the form in Eq.~(\ref{eq:el}), and for
each added level not among those in Table~\ref{tab:rydfreq}, 
a new adjusted variable $\delta_l$ is added.
The updated column vector of input data $Q_{\rm u}$, 
matrix of their covariances $V_{\rm u}$,
and column vector of variables $Z_{\rm u}$ are written in block form as
\begin{eqnarray}
 Q_{\rm u} = \left(\!\begin{array}{l} Q \\  Q_\delta \\ Q_E \end{array}\!\right)\!
;~
V_{\rm u} = \left(\!\begin{array}{ccc} V & T & 0 \\
                          T^\top &  S & 0 \\ 0 & 0 & V_E \end{array}\!\right)\!
;~
 Z_{\rm u} = \left(\!\begin{array}{c} Z \\  Z_\delta \end{array}\!\right)\! ,
\label{eq:blocks}
\end{eqnarray}
where $Q$, $V$,  and $Z$ are the corresponding sets of quantities used in the 2002
least-squares adjustment, $Q_\delta$ is the set of theoretical data
$\delta_l (\mbox{th}) =0$ for the new variables 
$\delta_l$, $Z_\delta$ is the new set of adjusted variables $\delta_l$, and $Q_E$
is input data for the new energy levels $E_l$.
In $V_{\rm u}$, where $V_{{\rm u}ik}={\rm cov}(Q_{{\rm u}i},Q_{{\rm u}k})$,
$S$ and $T$ are the sets of theoretical covariances
involving the new $\delta$s,
and $V_E$ is the set of covariances of the new levels $E_l$.
Since the input data for the new levels are unknown, we simply assume that
the uncertainties are very large and that there are no correlations among them
or with the rest of the input data.  
This yields the blocks of zeros in $V_{\rm u}$ and results in $V_E$ being diagonal.

The input data and adjusted variables are related by the set of
observational equations given by
\begin{eqnarray}
Q_{\rm u} \doteq F_{\rm u}(Z_{\rm u}); \qquad
\left(\!\begin{array}{l} Q \\  Q_\delta \\ Q_E \end{array}\!\right)
\doteq \left(\!\begin{array}{l} F(Z) \\  
Z_\delta \\ E(Z_{\rm u}) \end{array}\!\right) ,
\label{eq:qdefs}
\end{eqnarray}
where the dot over the equal sign
indicates that the equation represents the ideal relations between the
input data and the adjusted constants which are not simultaneously
satisfied, since the set of equations
is overdetermined.  In Eq.~(\ref{eq:qdefs}), $F$ is the set of 
functions in the observational equations of the 2002
adjustment, and $E$ is the set of expressions for the new energy levels
of the form in Eq.~(\ref{eq:el}).
The observational equations are linearized by writing the Taylor series
\begin{eqnarray}
Q_{\rm u} \doteq F_{\rm u}(Z_{\rm u}^{(0)}) 
+ A_{\rm u}(Z_{\rm u}-Z_{\rm u}^{(0)})
+ \cdots ,
\label{eq:expand}
\end{eqnarray}
where $A_{\rm u}$ is the matrix of derivatives
\begin{eqnarray}
A_{{\rm u}ij} = \frac{\partial F_{{\rm u}i}(Z_{\rm u}^{(0)})}
{\partial Z_{{\rm u}j}^{(0)}}~; \qquad
A_{\rm u} = \left(\begin{array}{cc} A & 0 \\ 0 & I \\ B & C\end{array}\right),
\label{eq:avup}
\end{eqnarray}
and neglecting higher-order terms.
In Eq.~(\ref{eq:avup}), $A$ is the matrix of derivatives from
the 2002 adjustment, $I$ is the identity matrix,
and $B$ and $C$ are derivatives of the new energy levels 
with respect to the old and new variables, respectively.
The truncated expression in Eq.~(\ref{eq:expand}) corresponds to
\begin{eqnarray}
Y_{\rm u} \doteq A_{\rm u} X_{\rm u} ,
\label{eq:yax}
\end{eqnarray}
where
$Y_{\rm u} = Q_{\rm u} - F_{\rm u}(Z_{\rm u}^{(0)})$ and
$X_{\rm u} = Z_{\rm u}-Z_{\rm u}^{(0)}$.

The update adjustment starts with
\begin{eqnarray}
Z_{\rm u}^{(0)} = \left(\begin{array}{c} \hat Z \\  0 \end{array}\right) ,
\end{eqnarray}
where $\hat Z$ is the final vector of constants from the 2002 
adjustment and
\begin{eqnarray}
 Y_{\rm u} = \left(\begin{array}{c} \hat Y \\  Y_\delta \\ 
Y_E\end{array}\right)
  = \left(\begin{array}{c} Q - F(\hat Z) \\  Q_\delta- Z_\delta^{(0)} \\ 
\! Q_E - E\big(Z_{\rm u}^{(0)} \big) \!\! \end{array}\right) ,
\end{eqnarray}
where $\hat Y$ is the final value of $Y$ from the 2002
adjustment and $Y_\delta = 0$.
The solution $\hat X_{\rm u}$ to Eq.~(\ref{eq:yax}), the value of $X_{\rm u}$ that
minimizes $(Y_{\rm u}-A_{\rm u}X_{\rm u})^\top V_{\rm u}^{-1}(Y_{\rm u}-A_{\rm
u}X_{\rm u})$,
is
\begin{eqnarray}
 {\hat X_{\rm u}} &=&  G_{\rm u}  A_{\rm u}^\top  V_{\rm u}^{-1}  Y_{\rm u}
; \quad
 G_{\rm u} = ( A_{\rm u}^\top  V_{\rm u}^{-1}  A_{\rm u})^{-1}.
\label{eq:upeqs}
\end{eqnarray}
The covariance matrix of the solution ${\hat X_{\rm u}}$ is $G_{\rm u}$,
and its calculation is the key to the update.
The Schur-Banachiewicz inverse formula 
\cite{schur1917,banach1937}
applied to the upper-left four
blocks of the matrix $V_{\rm u}$ in Eq.~(\ref{eq:blocks}) gives
\begin{eqnarray}
V_{\rm u}^{-1}\!\! &=&\!\! \left(\begin{array}{ccc} \!\!
V^{-1}\!\! + V^{-1}TRT^\top V^{-1} & 
-V^{-1}TR & 0 \\ 
-RT^\top V^{-1} & R & 0 \\
 0 & 0 & V_E^{-1} \!\!\!\!
\end{array}\right) , \qquad
\label{eq:vi}
\end{eqnarray}
where $R = (S-T^\top V^{-1}T)^{-1}$.
For increasing uncertainties of the unknown
input data for the new levels $E_l$, we have $V_E^{-1}\rightarrow0$, and
we work in this limit.  A direct calculation from Eqs.~(\ref{eq:avup}),
(\ref{eq:upeqs}), and (\ref{eq:vi}),  with $V_E^{-1}=0$, yields
\begin{eqnarray}
G_{\rm u}^{-1} &=& 
\left(\begin{array}{c@{\quad}c} 
G^{-1} + G^{-1}URU^\top G^{-1} & 
-G^{-1}UR \\ 
-RU^\top G^{-1} & R
\end{array}\right) , \qquad
\label{eq:gi}
\end{eqnarray}
where $G = (A^\top V^{-1} A)^{-1}$ is the matrix from the 2002 adjustment 
and $U = GA^\top V^{-1} T$.
Evidently, Eq.~(\ref{eq:gi}) is the 
Schur-Banachiewicz inverse expression for
\begin{eqnarray}
G_{\rm u} &=&  \left(\begin{array}{cc} G & U \\ U^\top & P
\end{array}\right) ,
\label{eq:g}
\\ \nonumber
\end{eqnarray}
provided $R = (P-U^\top G^{-1}U)^{-1}$, that is, if
\begin{eqnarray}
P &=& S - T^\top V^{-1}T + U^\top G^{-1}U
= S + DT,
\end{eqnarray}
where $D = T^\top V^{-1}\left(AGA^\top V^{-1}-I\right)$.
This result for $G_{\rm u}$ in terms of $G$ means that the 
exact result of the enlarged least-squares adjustment can be obtained
from results of the 2002 least-squares adjustment with
a relatively simple calculation.  That is, the matrix inversions needed for the
enlarged adjustment have effectively been carried out exactly, with the results
explicitly expressed in terms of the matrices and vectors of the 2002 
adjustment.  In particular,
\begin{eqnarray}
G_{\rm u}A_{\rm u}^\top V_{\rm u}^{-1} &=& 
\left(\begin{array}{ccc} GA^\top V^{-1} & 0 & ~~0 \\
  D  &  I  & ~~0
\end{array}\right) ,
\end{eqnarray}
so that
\begin{eqnarray}
\hat X_{\rm u} &=&
\left(\begin{array}{c} GA^\top V^{-1}\big[Q-F(\hat Z)\big] \\ 
 D\big[Q-F(\hat Z)\big] \end{array}\right) =
\left(\begin{array}{c} 0 \\ 
 D \hat Y \end{array}\right) ,
\label{eq:xuh}
\end{eqnarray}
or for the adjusted constants
\begin{eqnarray}
\hat Z_{\rm u}
= Z_{\rm u}^{(0)} + \hat X_{\rm u}
= \left(\begin{array}{c} \hat Z \\
 D \hat Y \end{array}\right)
\label{eq:zuh}
\end{eqnarray}
with covariance matrix ${\rm cov}(\hat Z_{\rm u}) = G_{\rm u}$.
More importantly, Eqs.~(\ref{eq:g}) and (\ref{eq:zuh}) show that both the values
and uncertainties of the new levels being calculated are influenced by their covariances
with the levels in the 2002 least-squares adjustment, 
while the values and uncertainties of the variables from that
adjustment are not changed at all.  Also, since the only adjusted
variables that change in the update appear linearly in Eq.~(\ref{eq:qdefs}), 
no iteration of the update is needed to reach the final result.

%%%%%%%%%%%%%%%%%%%%%%%%%%%%%%%%%%%%%%%%%%%%%%%%%%%%%%%%%%%%%%%%%%%%%
\def\vs{\vspace{0.0 pt}}
\def\vt{\vspace{1.8 pt}}

\begin{table}[t]
\caption{
Calculated transition frequencies 
in hydrogen and deuterium from the 1S state to
the 3S and 3D excited states.
}
\label{tab:gndfreq}
\begin{tabular}{c@{\quad}l@{\quad}l}
\toprule
\noalign{\vbox to 2 pt {}}
  Excited & \qquad Hydrogen & \qquad Deuterium \\
  state   & \qquad~  $\nu_{\rm H}$/kHz & \qquad~ $\nu_{\rm D}$/kHz  \\ 
\colrule
\noalign{\vbox to 2 pt {}}

  3S$_{1/2}$ \vs
 & $ 2\,922\,743\,278\,671.6(1.4)$ 
 & $ 2\,923\,538\,534\,391.8(1.4)$  \vt\\ 
 
   3D$_{3/2}$ \vs
 & $ 2\,922\,746\,208\,551.40(70)$ 
 & $ 2\,923\,541\,464\,741.75(72)$  \vt\\ 
 
   3D$_{5/2}$ \vs
 & $ 2\,922\,747\,291\,888.61(70)$ 
 & $ 2\,923\,542\,548\,374.66(72)$  \vt\\ 
 
 \botrule
 \end{tabular}
 \end{table}

%%%%%%%%%%%%%%%%%%%%%%%%%%%%%%%%%%%%%%%%%%%%%%%%%%%%%%%%%%%%%%%%%%%%%

\begin{table}[t]
\caption{
Examples of calculated transition frequencies 
in hydrogen and deuterium from the 2S state to
various S and D excited states.
}
\label{tab:exfreq}
\begin{tabular}{c@{\quad}l@{\quad}l}
\toprule
\noalign{\vbox to 2 pt {}}
  Excited & \qquad Hydrogen & \qquad Deuterium \\
  state   & \qquad~  $\nu_{\rm H}$/kHz & \qquad~ $\nu_{\rm D}$/kHz  \\ 
\colrule
\noalign{\vbox to 2 pt {}}

  3S$_{1/2}$ \vs
 & $ 456\,681\,865\,484.5(1.4)$ 
 & $ 456\,806\,126\,870.1(1.4)$  \\ 
 
   3D$_{3/2}$ \vs
 & $ 456\,684\,795\,364.30(69)$ 
 & $ 456\,809\,057\,220.01(69)$  \\ 
 
   3D$_{5/2}$ \vs
 & $ 456\,685\,878\,701.51(69)$ 
 & $ 456\,810\,140\,852.91(69)$  \vt\\
 
   4S$_{1/2}$ \vs
 & $ 616\,520\,150\,628.5(2.0)$ 
 & $ 616\,687\,903\,590.7(2.0)$  \\ 
 
   4D$_{3/2}$ \vs
 & $ 616\,521\,386\,393.3(1.7)$ 
 & $ 616\,689\,139\,553.8(1.7)$  \\ 
 
   4D$_{5/2}$ \vs
 & $ 616\,521\,843\,426.7(1.7)$ 
 & $ 616\,689\,596\,711.9(1.7)$  \vt\\ 

 \botrule
 \end{tabular}
 \end{table}
%%%%%%%%%%%%%%%%%%%%%%%%%%%%%%%%%%%%%%%%%%%%%%%%%%%%%%%%%%%%%%%%%%%%%

The energy levels and their covariances are thus given by
\begin{eqnarray}
\hat Q_E &=& E\big(\hat Z_{\rm u}\big)
\label{eq:ucov} \\ \nonumber
{\rm cov}(\hat Q_E) &=& BGB^\top + CU^\top B^\top 
+ BUC^\top + CPC^\top ,\quad
\end{eqnarray}
where the latter result is the lower-right block of the relation
${\rm cov}(\hat Q_{\rm u}) = A_{\rm u}G_{\rm u}A_{\rm u}^\top$.
The result from Eq.~(\ref{eq:ucov})
for a transition frequency $\nu_{lm}$ and its
standard uncertainty $u(\nu_{lm})$ for the transition
$l\rightarrow m$ is given by
\begin{eqnarray}
h \, \nu_{lm} &=& \hat Q_{El} - \hat Q_{Em}
\\ \nonumber 
h \, u(\nu_{lm}) &=& \left[u^2(\hat Q_{El}) 
- 2\,{\rm cov}(\hat Q_{El},\hat Q_{Em})
+ u^2(\hat Q_{Em})\right]^\frac{1}{2}
\end{eqnarray}
where $u^2(\hat Q_{Ei})
={\rm cov}(\hat Q_{Ei},\hat Q_{Ei})$, $i=l,m$.

Examples of
calculated transition frequencies in hydrogen and deuterium based on this 
update, starting from the results of the 2002 least-squares 
adjustment, are given in Tables~\ref{tab:gndfreq} and
\ref{tab:exfreq}.  Data from that adjustment needed for such a calculation
are available on the Web at physics.nist.gov/constants.
The frequencies in Tables ~\ref{tab:gndfreq} and
\ref{tab:exfreq} all have relative uncertainties 
that are smaller than the relative uncertainty of the Rydberg constant,
mainly as a result of the correlations between 
$R_\infty$, $R_{\rm p}$, and $R_{\rm d}$.
In some cases, these values are nearly five orders of magnitude more accurate than
the corresponding best previous values \cite{1977002}.
A database that gives the frequency of any transition between levels with
$n\le 200$ based on the calculations described here is maintained 
on the Web at physics.nist.gov/hdel.

Helpful conversations with G. W. Stewart are acknowledged by one of the 
authors (PJM).

%\bibliography{refs}

\end{document}